# Role of Joule Heating on Current Saturation and Transient Behavior of Graphene Transistors

Sharnali Islam, Zuanyi Li, Vincent E. Dorgan, Myung-Ho Bae, and Eric Pop, *Senior Member, IEEE*

*Abstract*—We use simulations to examine current saturation in sub-micron graphene transistors on $SiO_2$/Si. We find self-heating is partly responsible for current saturation (lower output conductance), but degrades current densities >1 mA/µm by up to 15%. Heating effects are reduced if the supporting insulator is thinned, or in shorter channel devices by partial heat sinking at the contacts. The transient behavior of such devices has thermal time constants of ~30-300 ns, dominated by the thickness of the supporting insulator and that of device capping layers (a behavior also expected in ultrathin body SOI transistors). The results shed important physical insight into the high-field and transient behavior of graphene transistors.

## I. INTRODUCTION

Graphene has attracted much interest for transistor applications, initially due to its high carrier mobility, ~$10^4$ $cm^2V^{-1}s^{-1}$ [1]. Recent work has also found drift velocity saturation at high field in graphene, at values several times higher than in silicon [2]. However, velocity saturation alone is not sufficient for current saturation, because the carrier density can continue to increase with drain voltage in a zero band gap material where the channel cannot be fully pinched off. Current saturation is important for low output conductance $g_o$ and amplifier gain [1, 3], and in practice it has been partly achieved through a combination of velocity saturation and electrostatic charge control [4, 5]. At the same time, high field transport in graphene is also influenced by self-heating [2, 6], as revealed by recent infrared and Raman thermal imaging [7-10].

In this work we examine the effect of self-heating on current saturation in sub-micron graphene-on-insulator (GOI) transistors through electro-thermal device simulations. We consider the role of the buried

Manuscript received XXX; revised YYY. This work was supported in part by the ONR, NRI and a NSF CAREER award. The review of this letter was arranged by Editor L. Selmi.

The authors are with the Dept. of Electrical & Computer Eng. and the Micro and Nanotechnology Lab, University of Illinois Urbana-Champaign (UIUC), Urbana, IL 61801, USA. Z. Li is also with the Physics Dept. at UIUC and M.-H. Bae is currently with the Korea Research Institute of Standards & Science, Daejeon 305-340, Korea (e-mail: epop@illinois.edu).

oxide thickness ($t_{box}$) under the graphene, and of the device length ($L$). We also observe that practical graphene devices can be operated in transient (pulsed) mode, and calculate their thermal time constants, i.e. the time scales over which the device temperature ramps up or cools down after electrical switching.

## II. CURRENT SATURATION

The schematic of a typical GOI transistor is shown in Fig. 1(a). Our simulations are based on the drift-diffusion approach, calculating carrier densities, electric field, drift velocity, potential, and temperature along the channel and contacts self-consistently. The simulator was extensively tuned against experimental data [8, 10], including contact effects [11]. The metal-graphene contact resistance per unit area used here is $\rho_C = 111 \ \Omega \cdot \mu m^2$ which is near the low end of the range for typical Pd- or Au-graphene contacts [11]. The Dirac voltage of simulated devices is $V_0 = 0$ V and the background temperature is $T_0 = 293$ K. Other parameters are as in Ref. [2], including compact models of mobility and velocity saturation dependent on carrier density and temperature. Since carrier mean free paths in typical GOI transistors are in the 20-80 nm range [8, 10], the model is most reliable for devices greater than ~0.1 μm.

We first investigate self-heating and current saturation in a device with channel length and width $L = W = 1$ μm. Fig. 1(b) shows the computed current vs. source-drain voltage ($I_D$-$V_{SD}$) of this GOI device on $t_{box} = 300$ nm $SiO_2$ with vertical electric fields of 0.3, 0.6, and 1.0 MV/cm, respectively. The dashed lines represent the current without self-heating ($T = T_0$), whereas the solid lines show some current degradation when Joule heating is self-consistently taken into account. Thus, the simulations suggest that self-heating is at least partially responsible for the current saturation observed in recent experiments on devices of comparable size and bias [4, 5].

Fig. 1(c) shows the total carrier density at the highest voltage and current bias point from Fig. 1(b), with and without self-heating. Interestingly, because graphene is a gapless material, we find that significant self-heating during operation can alter the *majority* carrier concentration by thermal carrier generation [2]. Thus, self-heating at high field influences not only the current saturation, but also the internal carrier distributions. Fig. 1(d) displays the temperature profiles corresponding to the maximum bias points for the three cases in Fig. 1(b). We note that sustained temperature rises $\Delta T > 200$ K have been linked with graphene device instability in experimental studies [2, 8].

We now study the peak temperature rise ($\Delta T$) and the percentage of saturation current degradation ($\Delta I/I_{sat}$) as we reduce $t_{box}$ from 300 nm to 50 nm. For all $t_{box}$, the peak $\Delta T$ and $I_{sat}$ are taken at the same $V_{SD} = 2$ V for vertical fields of 0.5, 1, and 2 MV/cm. Fig. 2(a) shows that the peak $\Delta T$ of devices with channel length $L = 1$ μm scales proportionally with $t_{box}$, as expected. However, we note that even in the limit of $t_{box} \rightarrow 0$ (graphene device directly on substrate, similar to graphene on SiC), the temperature rise is non-zero due to the thermal resistance of the graphene-substrate interface and that of heat spreading into the substrate itself [2, 10]. Fig. 2(b) shows $\Delta I/I_{sat}$ due to self-heating as a function of $t_{box}$. As a simple guideline, a ~5% degradation in $I_{sat}$ corresponds to $\Delta T \sim 170$ K above room temperature. For current density near ~1 mA/μm, as for the top curve in Fig. 1(b) on $t_{box} = 300$ nm, the current degradation due to Joule heating can be >10%, and for higher current densities the self-heating effect is proportionally larger. This can be partly compensated by reducing $t_{box}$ and $L$, as described here and below. In addition, elevated temperatures not only decrease device performance, but also have profound effects on long-term device and dielectric reliability [12].

We next explore the effect of Joule heating while scaling the channel length from 1 to 0.25 μm. Fig. 2(c) shows current-voltage curves computed with and without self-heating, indicating the self-heating effect is less in shorter channel devices. Fig. 2(d) also plots $\Delta I/I_{sat}$ due to self-heating vs. $t_{box}$ for the same channel lengths, at the same drain output conductance $g_o = \partial I_D/\partial V_{SD}$. Less current degradation at shorter channel lengths is explained by an enhanced role of heat dissipation "laterally" to the contacts in addition to "vertically" through the oxide. This was also recently observed in experimental work on sub-0.5 μm graphene nanoribbons (GNRs) [13] which noted that heat dissipation into the contacts begins to play a role when device dimensions become ≤ ~3 times the thermal healing length. The thermal healing length is a measure of the lateral heat diffusion along the graphene, $L_H \approx 0.2$ μm in graphene on SiO$_2$ and approximately half in GNRs which have lower thermal conductivity [8, 13]. Increased heat loss to the contacts is also seen as a sub-linear rise of current degradation in Fig. 2(d) for the shorter devices. Our present model numerically accounts for heat spreading into the substrate and the contacts [10], however this can also be treated to a good approximation analytically as in Ref. [13].

From a practical point of view, our simulations suggest that thermal effects are always significant in present graphene devices [4, 5] at lateral fields >1 V/μm. To avoid this, devices could be built on substrates with thinner $t_{box}$ or higher thermal conductivity (e.g. sapphire). However, some amount of self-heating can lead to better current saturation (lower $g_o$), but not carefully considering such effects can cause long-term device instability [12].

## III. THERMAL TRANSIENT

While the section above focused on effects of self-heating on DC characteristics, this section explores the transient device behavior. We perform finite element (FE) simulations as shown in Figs. 3(a-b); the device is symmetric about the cross-section marked by a dashed line in Fig. 1(a) and only one half needs to be simulated. Isothermal boundary conditions ($T = T_0$) are applied 10 μm away from the device at the bottom and right edges of the Si substrate, and other boundaries are adiabatic. We used temperature-dependent values for the thermal conductivity and heat capacity of the oxide [14], although the effect was relatively small, <5 %.

An input power of 0.5 mW is initially applied to the graphene channel, then turned off after 2.5 µs. Figs. 3(a-b) correspond to temperature distributions at the end of the heating pulse in a device without and with a capping layer (assumed to be SiO₂), respectively. These can be roughly understood as a typical device in a laboratory setup, vs. one that is integrated in a package. The temperature transient of the channel mid-point is shown in the inset of Fig. 3(c) for a capping layer $t_{cap}$ = 200 nm and $t_{box}$ = 250 nm. The thermal time constant $\tau$ is obtained by fitting the temperature decay as $T(t) = T_0 + T_1 e^{-t/\tau} + T_2(1 + t/\tau_0)^{-b}$, where $T_0$ = 293 K is the base temperature, $T_0 + T_1 + T_2$ is the steady-state peak temperature, and the third term is used to fit the long tail of the temperature decay due to the (small) residual heating transient of the Si substrate. Typical values for $b$ are in the range 0.5 to 2.5 and for $\tau_0$ from 10s to 100s of ns. In the "no cap" case $T_2$ is less than 30% $T_1$, but it becomes comparable to $T_1$ in cases "with cap".

The symbols in Fig. 3(c) summarize the calculated thermal time constant of the graphene device as the $t_{box}$ is scaled, for devices with a capping layer of 200 nm, 500 nm, and without ("no cap"). We can understand the scaling of the thermal time constant through a simple analytic model which includes each region as a lumped thermal resistance ($R_{th}$) and thermal capacitance ($C_{th}$). The thermal time constant $\tau$ is then the sum of contributions ($\Sigma R_{th} C_{th}$) from the relevant regions:

$$\tau \approx f_1 \frac{C_V}{k_{ox}} t_{box}^2 + \left[ f_2 \frac{C_V}{k_{ox}} t_{cap} + \frac{C_{Vm}}{k_m} t_m \right] \left( t_{box} + t_{eq} \right) \tag{1}$$

where $C_V$ = 1.76 MJ K⁻¹ m⁻³ and $C_{Vm}$ = 2.88 MJ K⁻¹ m⁻³ are heat capacities of the oxide and metal gate [15], $t_m$ is the thickness and $k_m$ (= 40 Wm⁻¹K⁻¹ for Pd) the thermal conductivity of the metal gate. The geometrical pre-factors $f_1 \sim 0.6$ and $f_2 \sim 0.8$ represent the fraction of the total temperature drop in the bottom oxide and top capping layer, respectively. The last term $t_{eq} \approx 200$ nm accounts for the thermal equivalent of transient cooling in the Si substrate (the limit $t_{box} \to 0$), consistent with previous studies on bulk CMOS devices [16]. We note the analytic model above could also be applied to other devices based on atomically-thin materials like MoS₂, or to ultra-thin body (UTB) silicon-on-insulator (SOI) transistors.

The model of Eq. (1) is plotted with dashed lines in Fig. 3(c), in good agreement with our FE simulations (symbols). The FE results are realistic within 10-20% accuracy, depending on the simulated domain size and choice of 3D vs. 2D simulations (the main trade-off being CPU time), however the main physical trends persist. These results suggest that thermal time constants follow an approximately quadratic dependence on $t_{box}$, which contributes to both the thermal resistance and thermal capacitance of the device.

The capping layer and metal gate do contribute to the term in Eq. (1) that is linear in $t_{box}$, but do not aid in "cooling" the device otherwise. Thus, a thicker gate or capping layer only add "thermal ballast" and can increase the thermal time constant. Interestingly, due to its thinness, the graphene layer itself does not influence the thermal transient of the device, which is dominated by heating of the surrounding materials. This is a unique aspect of devices based on graphene (or other 2D monolayer materials like $MoS_2$) vs. that of older silicon-on-insulator (SOI) technology, where a substantial thickness of the Si "body" retains a non-negligible heat capacity and thermal resistance [17, 18].

To conclude, we found that Joule heating during operation is partly responsible for current saturation and degradation observed in graphene device experiments. Self-heating is reduced with thinner dielectrics, and for sub-0.5 μm channel lengths the contacts begin to play a role in heat sinking. The thermal time constants of GOI devices are of the order ~100 ns, but strongly dependent on the materials surrounding the channel. Thermal transients are much slower than electrical transients (~1-10 ps), consistent with previous work on SOI technology [17, 18]. This implies that graphene devices are slow to heat up or cool down after electrical switching and, for instance, pulsed operation on time scales shorter than the thermal time constant can benefit from reduced self-heating compared to DC operating modes.

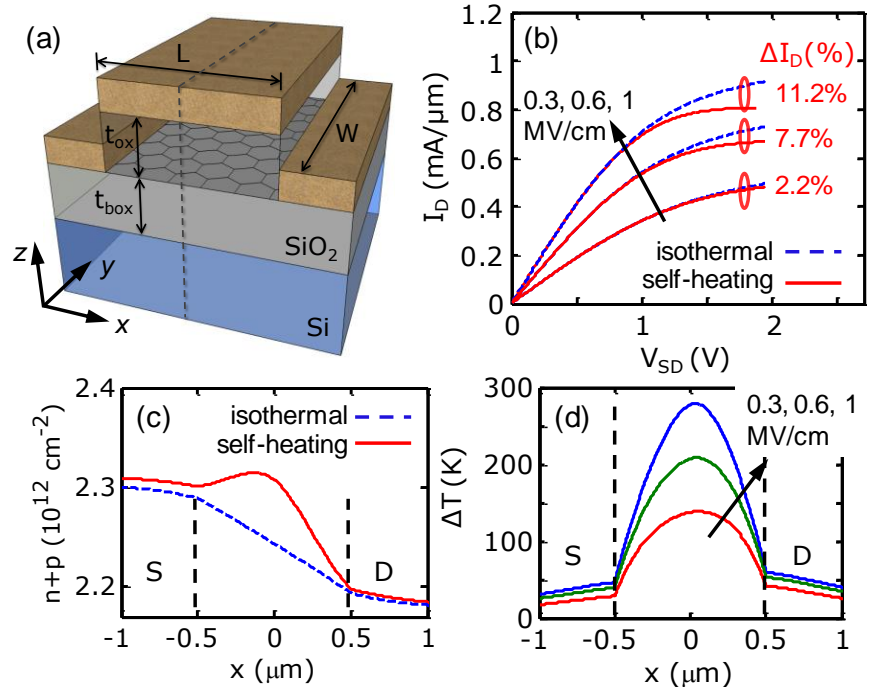

Fig. 1. (a) Schematic of simulated graphene device on SiO$_2$/Si substrate (image courtesy F. Lian). (b) Current saturation with self-heating (solid) compared to isothermal simulations (dashed) at three vertical E-fields (= $V_{GS}/t_{ox}$). (c) Carrier density along the channel at vertical field 1 MV/cm, with and without self-heating. (d) Temperature profiles at $V_{SD}$ = 2 V in (b) including self-heating. The device considered here has $L = W = 1$ μm and $t_{box} = 300$ nm.

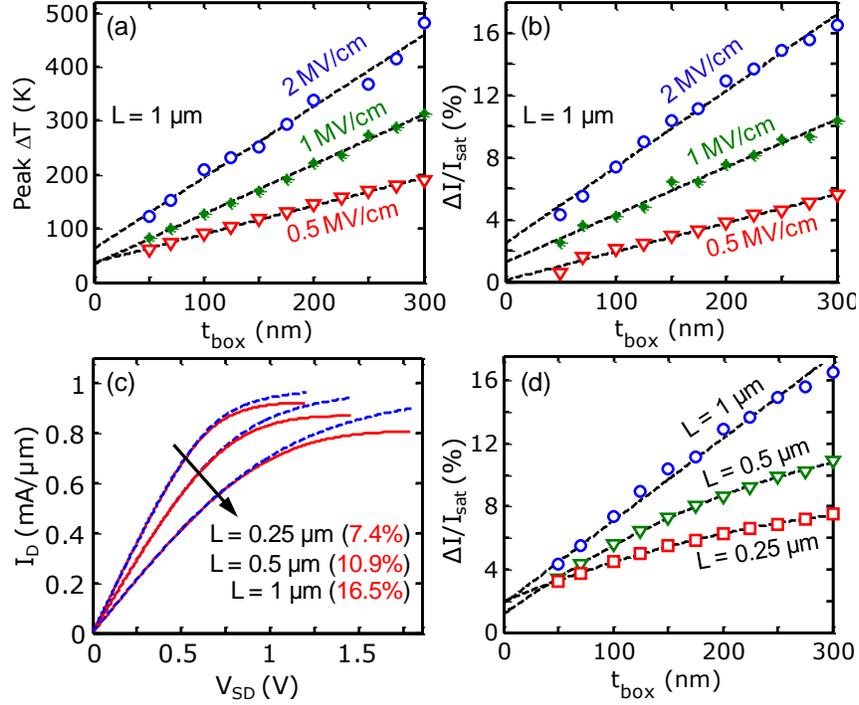

Fig. 2. (a) Calculated peak $\Delta T$ and (b) self-heating effect on saturation current as a function of $t_{box}$ for three vertical fields, at channel length $L = 1$ μm. Dashed lines are linear fits. (c) Current-voltage simulations with self-heating (solid) and without (dashed) for devices of $L = 0.25, 0.5, 1$ μm, on $t_{box} = 300$ nm and vertical field 2 MV/cm. (d) Self-heating effect on saturation current as a function of $t_{box}$ for the same three channel lengths and vertical field. Dashed lines show lower degradation and sub-linear dependence on $t_{box}$ for sub-0.5 μm channel lengths due to heat sinking effect of contacts.

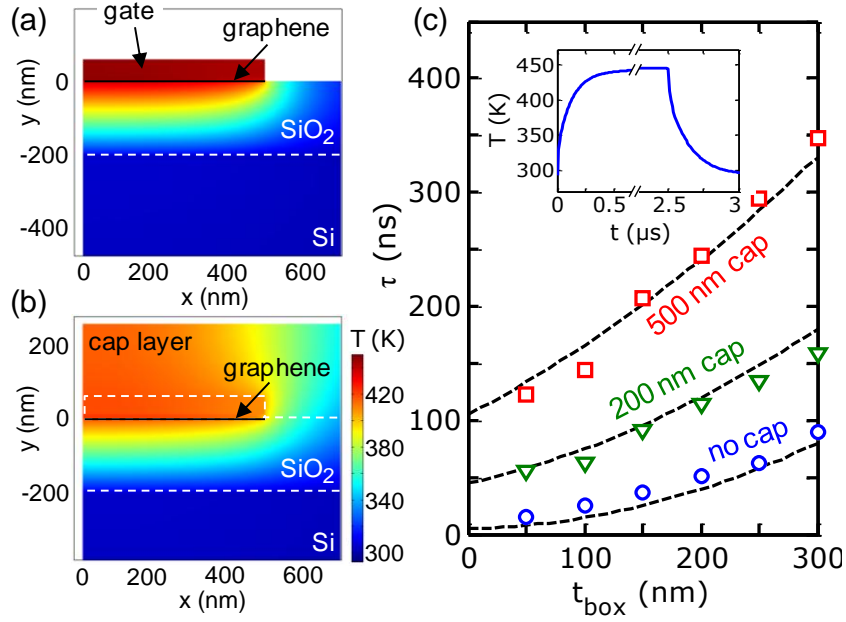

Fig. 3. Cross-section of graphene device temperature from Fig. 1 with (a) no capping layer and (b) 200 nm SiO₂ cap layer, 2.5 µs after a 0.5 mW input pulse. (c) Calculated thermal time constants of graphene devices as a function of $t_{box}$ without a capping layer (○), 200 nm cap layer (▽), and 500 nm cap layer (□). Dashed lines are fits with Eq. (1). The inset shows the temperature transient for $t_{cap}$ = 200 nm and $t_{box}$ = 250 nm. The power is turned on at $t$ = 0 s and off at $t$ = 2.5 µs.